\documentclass[useAMS,usenatbib]{mn2e}

\usepackage{graphicx}
\usepackage{color}

\def\ijl{$I_{ij\lambda}~$}
\def\ijlo{$(I_{ij\lambda})_O~$}

\title[PCA Tomography: how to extract information from datacubes]
{PCA Tomography: how to extract information from datacubes\thanks{Based on observations obtained at the Gemini Observatory, which is operated by the
Association of Universities for Research in Astronomy, Inc., under a cooperative agreement
with the NSF on behalf of the Gemini partnership: the National Science Foundation (United
States), the Science and Technology Facilities Council (United Kingdom), the
National Research Council (Canada), CONICYT (Chile), the Australian Research Council
(Australia), Minist\'erio da Ci\^encia e Tecnologia (Brazil) and SECYT (Argentina).}}
\author[Steiner et al.]{J. E. Steiner$^{1}$\thanks{E-mail:
steiner@astro.iag.usp.br (JES)}, R. B. Menezes$^{1}$, T.V. Ricci$^{1}$
and A. S. Oliveira$^{2}$\\
$^{1}$Instituto de Astronomia, Geof\'{\i}sica e Ci\^encias Atmosf\'ericas, Universidade de S\~ao Paulo, 05508-900,
    S\~ao Paulo, SP, Brasil\\
$^{2}$IP\&D, Universidade do Vale do Para\'{\i}ba, Av. Shishima Hifumi, 2911, CEP 12244-000, S\~ao Jos\'e dos Campos, SP, Brasil}
\begin{document}

\date{Accepted . Received }

\pagerange{\pageref{firstpage}--\pageref{lastpage}} \pubyear{2008}

\maketitle

\label{firstpage}

\begin{abstract}
Astronomy has evolved almost exclusively by the use of spectroscopic and imaging techniques, operated separately. With the development of modern technologies it is possible to obtain datacubes in which one combines both techniques simultaneously, producing images with spectral resolution. To extract information from them can be quite complex, and hence the development of new methods of data analysis is desirable.

We present a method of analysis of datacube (data from single field observations, containing two spatial and one spectral dimension) that uses PCA (Principal Component Analysis) to express the data in the form of reduced dimensionality, facilitating efficient information extraction from very large data sets. PCA transforms the system of correlated coordinates into a system of uncorrelated coordinates ordered by principal components of decreasing variance. The new coordinates are referred to as eigenvectors, and the projections of the data onto these coordinates produce images we will call tomograms. The association of the tomograms (images) to eigenvectors (spectra) is important for the interpretation of both.  The eigenvectors are mutually orthogonal and this information is fundamental for their handling and interpretation. When the datacube shows objects that present uncorrelated physical phenomena, the eigenvector's orthogonality may be instrumental in separating and identifying them. By handling eigenvectors and tomograms one can enhance features, extract noise, compress data, extract spectra, etc.

We applied the method, for illustration purpose only, to the central region of the LINER galaxy NGC 4736, and demonstrate that it has a type 1 active nucleus, not known before. Furthermore we show that it is displaced from the centre of its stellar bulge.
\end{abstract}

\section{Introduction}

Throughout the $20^{th}$ Century, astronomy has developed through the use of imaging and spectroscopic techniques, analysed independently. Extracting information from these types of data, requires relatively simple tools. With the advent of panoramic spectroscopic devices such as Integral field units -- IFUs -- and Fabry-Perot spectrographs, it is possible to construct datacubes of immense proportions that present data in three dimensions: two spatial and one spectral. The analysis of these data may become complex and overwhelming, as it may involve tens of millions of pixels. More concerning is that, given this complexity, only some restricted subset of the data ends up being analysed (kinematical maps, line flux ratios, extinction and excitations maps, etc.); the rest is at the risk of being largely ignored. New techniques that allow us to extract information in a condensed, fast and optimized form are therefore necessary and welcome.

In this paper we present a method of datacube interrogation that uses Principal Component Analysis -- PCA. This method condenses the significant information content associated with the data, through effective dimensional reduction, facilitating its interpretation. PCA compresses the data expressed as a large set of correlated variables in a small but optimal set of uncorrelated variables, ordered by their principal components. Clearly our shared goal of analysing data is to extract physical information from them; a dimensional reduction does not necessarily produce valuable information, but an appropriate choice of coordinates may help. PCA is a non-parametric analysis. This means that there are no parameters or coefficients to adjust that somehow depend on the user’s experience and skills, or on physical and geometrical parameters of a proposed model. PCA provides a unique and objective answer. In the traditional scientific method one formulates questions and looks to the data for answers. In this new strategy, PCA produces the answer; the user's challenge is to interpret the results. This process is not always difficult, but often plain  of subtleties.

PCA has been used many times in the astronomical literature. For instance, \citet{deeming} used this technique to analyse and classify stellar spectra; this approach was improved by \citet{whitney}. Applications to modern stellar spectroscopy can be found in \citet{bailer} and \citet{re}. The technique was also used for morphological \citep{lahav} and spectral \citep{sodre} classification of galaxies and QSOs \citep{boroson}. Images of supernova remnants have been analysed with PCA technique \citep{warren}. A more extended presentation of this technique is given in \citet{murtag} and \citet{fukunaga}.

Most of the applications of PCA in astronomy are related to find eigenvectors across a population of objects. In the present case we want to apply the technique to a single datacube in which the objects are spatial pixels of an individual field, containing a single galaxy, nebula or a set of stars. We identify eigenvectors (the uncorrelated variables) that we refer to as eigenspectra, and tomograms, which are images of the data projected in the space of the eigenvectors. In traditional tomographic techniques one obtains images that represent ``slices'' in tridimensional space (the human body, for example) or in velocity space (Doppler Tomography). In PCA Tomography one obtains images that represent ``slices'' of the data in the eigenvectors space (tomograms). The good news is that each tomogram has associated with it an eigenspectrum. The simultaneous analysis of the eigenspectra and associated tomograms brings a new perspective to the interpretation of both.

With the aim of illustrating the PCA Tomography method, we have applied it to a Gemini GMOS--IFU datacube of the nuclear region of the nearby LINER galaxy NGC 4736 (M94). The LINER characteristics of NGC 4736 are considered to be related to an atypical population of stars, as it is an aging starburst galaxy \citep{eracleous, cid}. Applying the PCA methodology, we show that it has a \textit{bona fide} type 1 Active Galactic Nucleus -- AGN -- displaced from the centre of its stellar bulge.

\section{From a datacube to a data matrix}

Our aim is to analyse datacubes in which we have two spatial and one spectral dimension. Each pixel of this original three-dimensional datacube has intensity \ijlo; here $i$ and $j$ define a spatial pixel and $\lambda$ a spectral pixel. We will assume that the datacube has $n = \mu \times \nu$ spatial pixels and $m$ spectral pixels.
The mean intensity of all spatial pixels for a given $\lambda$ is

\begin{equation}
\label{qlambda}
Q_\lambda = \frac{1}{n} \sum_{i=1}^{\mu} \sum_{j=1}^{\nu} (I_{ij\lambda})_O
\end{equation}

\noindent $Q_\lambda$ being the average spectrum of the datacube. The intensity adjusted to the mean is

\begin{equation}
\label{minusmean}
I_{ij\lambda} = (I_{ij\lambda})_O - Q_\lambda
\end{equation}

It is important to note at this point that all emission with null variance across the spatial pixels (for a given wavelength or spectral energy) are incorporated into the mean and subtracted out. This is the case, for instance, for the sky emission that is constant over the field of view (FoV).

Now we organize the new datacube \ijl (which has zero mean) into a matrix $\mathbfss{I}_{\bmath{\beta\lambda}}$ of $n$ rows (spatial pixels, referred to here as objects) and $m$ columns (spectral pixels, referred to here as properties). Then $\beta$ can be expressed as

\begin{equation}
\label{beta}
\beta = \mu(i-1)+j
\end{equation}

The datacube transformed into the matrix $\mathbfss{I}_{\bmath{\beta\lambda}}$ will be the subject of the PCA Tomography method.

\section{Elements of Principal Component Analysis -- PCA}

Principal Component Analysis (PCA) is a technique used to analyse multidimensional datasets. Its is a quite efficient method to extract information from a large set of data as it allows us to identify patterns and correlations in the data that in other ways would hardly be noticed. Mathematically it is defined as a linear orthogonal transformation that expresses the data in a new (uncorrelated) coordinate system such that the first of these new coordinates, $E_1$, (eigenvector 1) contains the largest variance fraction, the second variable, $E_2$, contains the second largest variance fraction and so on. These new coordinates generated by the PCA are, by construction, orthogonal to one another.
For a more detailed description of the PCA method, see \citet{murtag}, \citet{fukunaga}, \citet{johnson} and \citet{hair}.

In many PCA implementations normalization is done, so that variance is uniform (and generally unity) within the data. We will not adopt this strategy as we are interested in retaining the relative spectral line intensities. Therefore we will analyse the covariance matrix and not the correlation matrix.

The covariance matrix of $\mathbfss{I}_{\bmath{\beta\lambda}}$ can be expressed as 

\begin{equation}
\mathbfss{C}_{\bf{cov}} = \frac{[\mathbfss{I}_{\bmath{\beta\lambda}}]^T \bmath{\cdot} \mathbfss{I}_{\bmath{\beta\lambda}}} {n-1}
\end{equation}

The matrix $\mathbfss{C}_{\bf{cov}}$ is square and has $m$ rows and columns (equal to the number of the original spectral pixels). The covariance matrix has some relevant properties. One is that it is symmetric, 

\begin{equation}
\mathbfss{C}_{\bf{cov}} = [\mathbfss{C}_{\bf{cov}}]^T
\end{equation}

The main diagonal elements correspond to the variances of each of the isolated variables, while the other (cross) elements correspond to the covariance between two distinct properties. The $m \times m$ covariance matrix has $m$ eigenvectors, $E_k$, each one associated with one eigenvalue, $\Lambda_k$. $E_{k}$ are the new uncorrelated coordinates and $k$ is the order of the eigenvector that can vary from 1 to $m$; the eigenvectors are ordered by decreasing value of each associated $\Lambda_k$, which is the variance of each component, to form the characteristic matrix, $\mathbfss{E}_{\bmath{\lambda k}}$, in which columns correspond to eigenvectors. Note that, in order for all eigenvectors to be defined, we require that $n\geq m$.

The transformation that corresponds to the PCA can be represented by the following formula:

\begin{equation}
\label{tkb}
\mathbfss{T}_{\bmath{\beta k}} = \mathbfss{I}_{\bmath{\beta\lambda}} \bmath{\cdot} \mathbfss{E}_{\bmath{\lambda k}}
\end{equation}

\noindent where $\mathbfss{T}_{\bmath{\beta k}}$ is the matrix containing the data in the new coordinate system.

As the aim of  PCA is to express the original data on the new system of uncorrelated coordinates, one concludes that the ideal covariance matrix of the data in this new coordinate system ($\mathbfss{D}_{\bf{cov}}$) must be diagonal, that is, the covariance between the coordinates must be zero.  One may say that the PCA execution consists in determining the matrix $\mathbfss{E}_{\bmath{\lambda k}}$ that satisfies equation~\ref{tkb} and so that $\mathbfss{D}_{\bf{cov}}$ is diagonal:

\begin{equation}
\mathbfss{D}_{\bf{cov}} = \frac{[\mathbfss{T}_{\bmath{\beta k}}]^T \bmath{\cdot} \mathbfss{T}_{\bmath{\beta k}}}{n-1}
\end{equation}

The diagonal elements of $\mathbfss{D}_{\bf{cov}}$ are the eigenvalues.

\section{Eigenspectra and Tomograms}

In the case of datacubes of astronomical interest, it is usual to have two-dimensional images with spectra associated with each spatial pixel. In calculating the PCA of such datacubes, one obtains eigenvectors as a function of wavelength, energy or frequency (properties), that we will also refer to as eigenspectra.

On the other hand, $\mathbfss{T}_{\bmath{\beta k}}$ represents data in a new coordinate system. As our objects are spatial pixels, their projection onto a given eigenvector may be represented as a spatial image. Each column of $\mathbfss{T}_{\bmath{\beta k}}$ can now be transformed into a two-dimensional image, $T_{ijk}$, using equation~\ref{beta}. We will refer to these images $T_{ijk}$ as \textbf{tomograms}, since they represent ``slices'' of the data in the space of the eigenvectors. 

When a stellar-like feature is present within the FoV, contiguous pixels tend to be correlated as the signal is influenced by the spatial Point Spread Function (PSF). Real structures have, thus, a minimum scale given by the PSF, usually determined by the seeing or intrinsic spatial instrumental resolution.

Analysing tomograms simultaneously with eigenspectra brings together a wealth of information. Spectral characteristics may be identified with features in the image and \textit{vice-versa}. Interpreting such associations facilitates the understanding of the three-dimensional structure within the datacube. In section~\ref{aplic} we will see an application of this and its potential will become clear.

\section{Reconstruction, compression, cosmetics and flux calibration of datacubes}

\begin{figure}
\centerline{\includegraphics[width=0.5\textwidth]{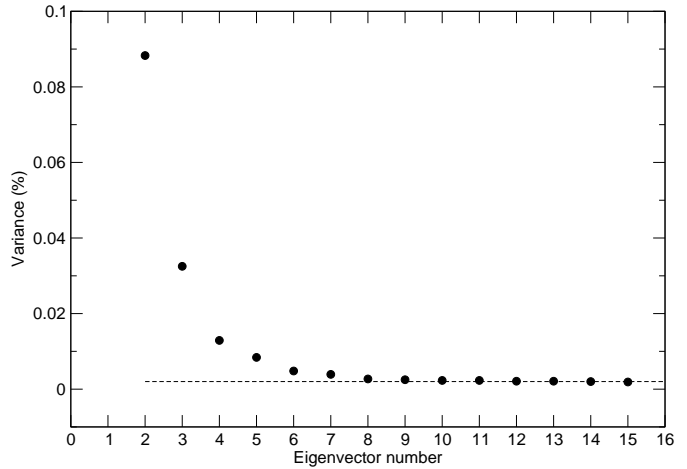}}
\caption{The ``scree test'' applied to the first 15 eigenvalues of the NGC 4736 datacube. One can see that the graph levels off for eigenvetors above 7. Eigenvalue 1 is out of scale.  \label{fig1}}
\end{figure}

It is, of course, possible to reconstruct the original datacube from all the eigenvectors and tomograms. It is also, however, possible to partially reconstruct the datacube using only those eigenvectors and tomograms that contain interesting or relevant information, ignoring those that contain noise. It is not straightforward to know where the signal stops and the noise becomes dominant. The Kayser criterion \citep{johnson} suggests that the limit is the mean eigenvalue. This criterion seems to select too few eigenvectors. One can, else, use the ``scree test'' \citep{hair} which  is illustrated in Fig.~\ref{fig1}. In practice, the number of relevant eigenvectors depends on the number of uncorrelated physical phenomena represented in the object. There is no way to know \textit{a priori}; each case must be examined by the user and the actual delineations to be considered depends on his skills and predilections. Let us reconstruct the datacube taking as a characteristic matrix the set of all eigenvectors that have relevance until $k=r$, ignoring all others. In this case the reconstructed matrix $\mathbfss{I'}_{\bmath{\beta\lambda}}(\leq r)$ is

\begin{equation}
\label{ilinha}
\mathbfss{I'}_{\bmath{\beta\lambda}}(\leq r) = \mathbfss{T}_{\bmath{\beta k}}(\leq r) \bmath{\cdot} [\mathbfss{E}_{\bmath{\lambda k}}(\leq r)]^T
\end{equation}

\noindent where $\mathbfss{E}_{\bmath{\lambda k}}(\leq r)$ is the characteristic matrix with columns corresponding to eigenvectors until $k=r$ and
 $\mathbfss{T}_{\bmath{\beta k}}(\leq r)$ is the data matrix in the new coordinate system containing eigenvectors only to $k=r$. From the matrix
 $\mathbfss{I'}_{\bmath{\beta\lambda}}(\leq r)$ one can reconstruct the datacube $I'_{ij\lambda}(\leq r)$. The datacube $I'_{ij\lambda}(\leq r)$ 
contains many more data (pixels) than do $\mathbfss{E}_{\bmath{\lambda k}}(\leq r)$ and 
$\mathbfss{T}_{\bmath{\beta k}}(\leq r)$, even if it does not contain more information. Note that for data transmission, it is much faster to send $\mathbfss{E}_{\bmath{\lambda k}}(\leq r)$ and
$\mathbfss{T}_{\bmath{\beta k}}(\leq r)$ than $I'_{ij\lambda}(\leq r)$, which can be reconstructed using equation~\ref{ilinha}. This form of data compression has practical applications, for example in data transmission.

Remembering that the average spectrum, $Q_\lambda$, was subtracted from the original data (see equation~\ref{minusmean}), it can now be added to the reconstructed datacube, to recover calibration:

\begin{equation}
\label{somaQlambda}
(I'_{ij\lambda}(\leq r))_O = I'_{ij\lambda}(\leq r) + Q_\lambda
\end{equation}

In this case, the reconstructed datacube does not have the the variance (presumably mostly noise) contained in eigenvectors $r<k<m$. 

Recall now that the eigenvalue $\Lambda_k$ can be expressed as

\begin{equation}
 \Lambda_k = \frac{[\mathbfss{T}_{\bmath{\beta k}}(k)]^T \bmath{\cdot} \mathbfss{T}_{\bmath{\beta k}}(k)} {n-1} 
\end{equation}

\noindent where $\mathbfss{T}_{\bmath{\beta k}}(k)$ is the matrix containing only the column corresponding to the projection of the data on $E_k$.
The sum of the variance contained in eigenvectors $r<k<m$, or ``noise''(this sum could still contain also some ``signal'') may, thus, be evaluated as $\sigma$, in the rms sense, between images ($I'_{ij\lambda}(\leq r))_O$ and $(I_{ij\lambda})_O$ and may be expressed as

\begin{equation}
\sigma^2 = \sum_{k=r+1}^{k=m} \Lambda_k
\end{equation}

One may also reconstruct the datacube of a single eigenvector $E_k$. In this case, 

\begin{equation}
\label{11}
\mathbfss{I'}_{\bmath{\beta\lambda}}(k) =   \mathbfss{T}_{\bmath{\beta k}}(k) \bmath{\cdot} [\mathbfss{E}_{\bmath{\lambda k}}(k)]^T
\end{equation}

\noindent where $\mathbfss{E}_{\bmath{\lambda k}}(k)$ is the matrix containing the column corresponding to the eigenvector $E_k$.
 From the matrix $\mathbfss{I'}_{\bmath{\beta\lambda}}(k)$, one can reconstruct the datacube $I'_{ij\lambda}(k)$, that
 presents the original dimensions but contains information from the object-eigenvector $E_k$ only.

It is quite common to have cosmetic problems in the datacubes. This can happen, for example, as incomplete removal of cosmic rays and hot/cold pixels. In this situation the ``defect'' may appear as a specific eigenvector or correlated with some set of other properties. It is usually easily detected and can be removed in the appropriate eigenvector, by excluding or by correcting it. 

Flux calibration of a reconstructed datacube can be done by adding the average spectrum, $Q_\lambda$ (as shown in equation~\ref{somaQlambda}). In general this process can only be done if the datacube is reconstructed with all its components. If we want to ignore the noise, then we are incorporating a small often negligible error. If the average spectrum has two components, say a  stellar and a line emitting component, then one could, in principle, separate the two and, by separating them in the reconstructed datacube, calibrate both. The final result is additive. In section~\ref{aplic} we will see an application of such a procedure.

When the datacube is reconstructed, it may have a spatially defined field without any object, representing background only; because the initial average spectrum was subtracted, this field should present a negative signal. One way to fix this is by adding the average spectrum, as seen above. But, sometimes, this is not desirable, for example, when the average spectrum contains sky emission. In this case it might be better to calculate the average spectrum from the background (this average spectrum is not affected by sky emission, because it was obtained from eigenvectors), multiply it by $-1$ and add it to the entire cube. This way we ensure that the spectrum of the background is reset to zero. This procedure may be particularly useful for sky subtraction in data obtained with Fabry-Perot.

\section{Feature suppression and enhancement} \label{enhancement}

In order to suppress or emphasize the properties of a given feature ``A'' (defined by its image or spectral characteristics), we construct a feature factor $\Gamma_k(A)$, for each eigenvector $E_k$, such that

\begin{equation}
\Gamma_k(A) = 1; 0
\end{equation}

\noindent depending on whether eigenvector $k$ is to be suppressed (0) or not (1) - and this is a user-chosen value. 
Feature ``A'' may be a star, a galactic nucleus or, else, a spectral class or a feature such as the Broad Line Region -- BLR -- of an Active Galactic Nucleus. With such a strategy we can reconstruct a datacube in which the desired feature is suppressed or enhanced:

\begin{equation}
\label{13}
I'_{ij\lambda}(A) = \sum_{k} [I'_{ij\lambda}(k) \bmath{\cdot} \Gamma_k(A)]
\end{equation}

\noindent or obtaining directly

\begin{equation}
\label{14}
\mathbfss{I'}_{\bmath{\beta\lambda}}(A) =   \mathbfss{T}_{\bmath{\beta k}} \bmath{\cdot}  [(\mathbfss{E}_{\bmath{\lambda k}})_\Gamma]^T
\end{equation}

\noindent where $(\mathbfss{E}_{\bmath{\lambda k}})_\Gamma$ corresponds to the matrix $\mathbfss{E}_{\bmath{\lambda k}}$ that had each of its columns (that correspond to each of the eigenvectors $E_k$) multiplied by the corresponding $\Gamma_k$ factors. From the matrix $\mathbfss{I'}_{\bmath{\beta\lambda}}(A)$ one can reconstruct the corresponding datacube $I'_{ij\lambda}(A)$. Now one can project the intensities on the dimensions $ij$, showing explicitly the image with the suppression or enhancement object ``A''. The spectrum of the enhanced object could also be extracted; such an example is shown in Fig.~\ref{fig4}.

We can follow this with an alternative or parallel strategy. Instead of adding the intensities (as in equation~\ref{13}), one can add the intensity associated to each eigenvector divided by its variance. To do this, we first multiply each column of the matrix $\mathbfss{T}_{\bmath{\beta k}}$ (which correspond to each of the tomograms) by the factor $N_k$, given by

\begin{equation}
\label{15}
N_k = \frac{1}{(\Lambda_k)^{\frac{1}{2}} \cdot (n-1)}
\end{equation}

\noindent where $n$ is the number of spatial pixels in the image. $N_k$ corresponds to a normalization factor. A consequence of the factor $(n-1)$ is that the sum of the square of all spatial pixels is 1. Then we define 

\begin{equation}
\label{16}
\mathbfss{V'}_{\bmath{\beta\lambda}}(A) = (\mathbfss{T}_{\bmath{\beta k}})_N  \bmath{\cdot} [(\mathbfss{E}_{\bmath{\lambda k}})_\Gamma]^T 
\end{equation}

\noindent where $(\mathbfss{T}_{\bmath{\beta k}})_N$  corresponds to the matrix $\mathbfss{T}_{\bmath{\beta k}}$ with unit variance, that is, with each of its columns (that correspond to each tomograms) multiplied by the corresponding $N_k$ factors. It is important to note that each tomogram has zero mean. From the matrix
$\mathbfss{V'}_{\bmath{\beta\lambda}}(A)$ one can reconstruct the corresponding datacube $V'_{ij\lambda}(A)$. The difference is that in the case of $I'_{ij\lambda}$ one emphasizes the intensity component of each eigenvector while in the case of $V'_{ij\lambda}$ one has the distinct characteristic of all eigenvectors with the same weight. The advantage of $V'_{ij\lambda}$ with respect to $I'_{ij\lambda}$ is that the former shows more ``colorful'' features, enhancing the many characteristics of all eigenvectors,  but it may also enhance the noise as it gives similar weight to all isolated eigenvectors. Fig.~\ref{fig5} shows such an example.

\section{Application: The central region of the LINER galaxy NGC 4736}\label{aplic}

Let us illustrate the application of the method of PCA Tomography to a particular case. We will attempt to answer the following question: is there a supermassive black hole in the nearby LINER galaxy NGC 4736? LINERs are a class of objects with diverse nature \citep{heckman}. Although most of them seem to host an AGN in the sense that they are powered by accretion onto a supermassive black hole, some objects have not shown any evidence of this. NGC 4736 is somewhat peculiar because it presents a stellar population that corresponds to an aging starburst. Could this explain its LINER nature? See \citet{eracleous} and \citet{cid} for a more detailed discussion.

\begin{figure}
\centerline{\includegraphics[width=0.4\textwidth]{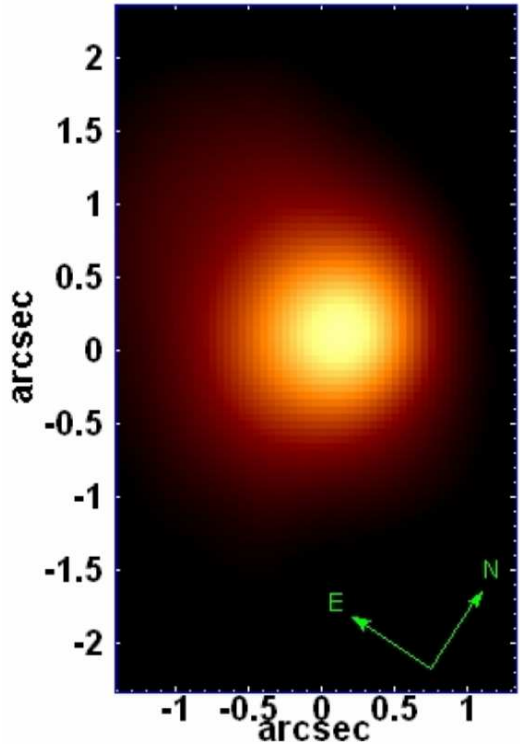}}
\caption{The projection of the average intensities $<(I_{ij})_O>$ for each $\lambda$  of the original datacube onto the spatial pixels ij. The bulge has a comet-like morphology. \label{fig2}}
\end{figure}

\subsection{The data}

In an attempt to solve the puzzle associated with this galaxy, we observed it with the Gemini Multi Object Spectrograph (GMOS -- \citet{hook}, \citet{allington}), operated in the Integral Field Unit (IFU) mode. The data were obtained on 2006 June 23 with the Gemini North Telescope. The datacube was obtained using 500 fibers on the object and 250 fibers on sky, 1 arcminute away. The spectral resolution was R=2900 covering from 4700 to 6800 {\AA}. Three 20 minute integration were obtained.

\begin{figure*}
\vspace{1.5cm}
\centerline{\includegraphics[width=0.7\textwidth]{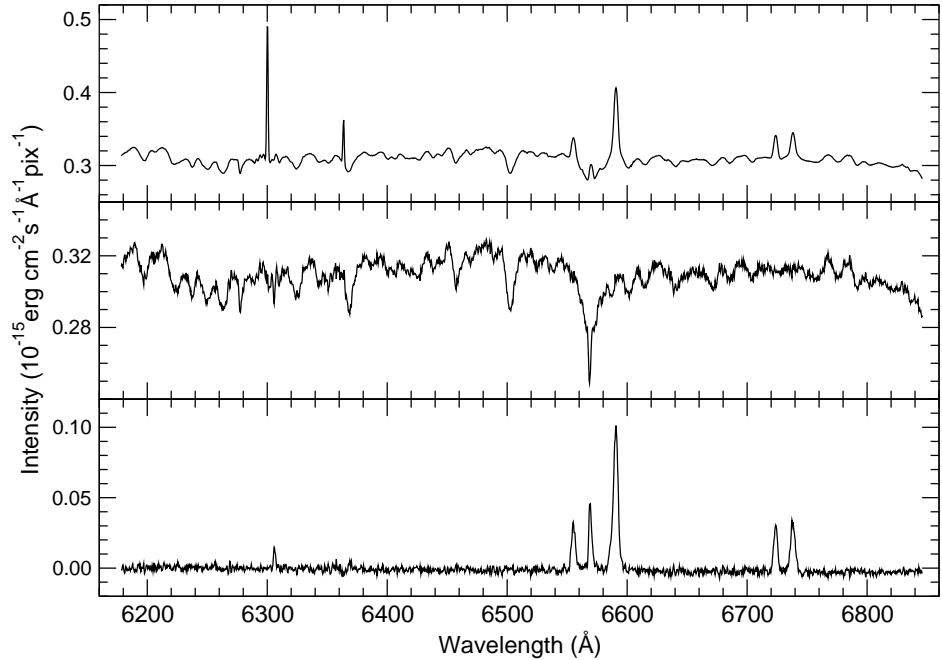}}
\caption{The average spectrum (\textit{top diagram}), as defined in equation~\ref{qlambda}; the stellar (\textit{middle diagram}) and gaseous (\textit{lower diagram}) components of the average spectrum were obtained by subtracting a scaled emission line (could also have been a stellar) template. See sections~\ref{stellargas} and \ref{flux} for how this is done. \label{fig3}}
\end{figure*}

The sky fibers actually observed the inner ring of the galaxy, as there was no other way to position them. For this reason these sky observations were not used and the datacube we analyzed did not have any kind of sky subtraction. In this situation PCA analysis is still possible as sky has no spatial variance and is incorporated in the average spectrum, being removed from the cube right in the beginning. Two strong telluric emission lines can be seen in the average spectrum (Fig.~\ref{fig3}) and present no sign in any of the eigenvectors or reconstructed $I'_{ij\lambda}$ or $V'_{ij\lambda}$ cubes. Only when dealing with flux calibration some special care must be taken, and in sections~\ref{stellargas} and \ref{flux} we show that this is still possible.

Comparison CuAr lamps, flatfields, twilight flats and bias images were taken to reduce and calibrate the data. The data reduction was done with the IRAF package using the $gemini.gmos$ task package that handles the bias and background subtraction, cosmic ray rejection, CCD and fiber sensitivity correction, wavelength and flux calibration and construction of the datacubes. Our final scientific datacube was extracted with a spatial oversampling of 0.05 arcsec pixel$^{-1}$ ($4\times4$ data pixel per fiber) compared to the real spatial resolution as determined by the 0.55 arcsec seeing experienced at the time of observations.
The datacube has 6,200 spectral pixels, with 0.34 {\AA} pixel$^{-1}$ spectral sampling.

As the GMOS atmospheric dispersion corrector was not operational, the differential atmospheric refraction was appreciable, giving wavelength distortions throughout the datacube. To evaluate this, we used the formula from \citet{filippenko} and applied our own algorithm for differential atmospheric refraction correction \citep{steiner}. This algorithm corrects each pixel for the atmospheric differential refraction to an accuracy of about 1/20 of the seeing disk.

A Richardson-Lucy deconvolution algorithm \citep{richardson, lucy} was applied to all images in the datacube using 6 iterations. This procedure has two effects: it sharpens the PSF while suppressing the high frequency noise. If the number of iterations is too small, these improvements are negligible while, if the number is too large, low frequency noise is introduced. We found that $\sim 6$ iterations was a good compromise; the delivered PSF after deconvolution reduced the FWHM of the PSF by a factor of 1.4. The adopted instrumental PSF for the purposes of the deconvolution was gauged from the spatially compressed image as a Gaussian having a FWHM of 0.47 arcsec. This type of deconvolution is compatible with PCA; we have experienced these procedures with dozens of cases, involving datacubes of galaxies, nebulae and stars, with good results.

In this paper we will analyse only the data corresponding to the GMOS red CCD, with a wavelength range from 6179 to 6848 {\AA} and 1976 spectral
and 5170 spatial pixels, after trimming some of the borders because of the atmospheric differential refraction correction. The full data set is analysed in \citet{steiner}.

\subsection{Eigenvectors, Tomograms and Eigenvalues}

We are ready to perform the PCA Tomographic analysis of the datacube. Before doing so, we have subtracted the average of all spatial pixels, $Q_\lambda$, for each wavelength pixel. The original datacube has its spatial projection shown in Fig.~\ref{fig2} while the average spectrum (see equation~\ref{qlambda}) is shown in the top diagram of Fig.~\ref{fig3}.
	
How many eigenvectors do we want to work with? Applying the scree test (see Fig.~\ref{fig1}) one can see that the relevant eigenvectors are limited to the first seven. It is always good to examine the eigenvectors/tomograms, case by case; in the present example the features associated to gas emission seem to be present in the first seven, disappearing in the noise for eigenvectors of higher order. This confirms the conclusion from the scree test. However, this does not mean the information about the stellar population is not encoded in eigenvectors of higher order. In dealing with the stellar component, one should keep this in mind. In what follows we are interested in the features associated to the emission lines and will limit ourselves to the first eight eigenvectors.

The eight principal components are shown as eigenvectors and tomograms in Appendix A and their eigenvalues, in Table~\ref{eigenvalue}. As can be seen, eigenvector 1 contributes 99.74 per cent of the variance. This means that this eigenspectrum basically replicates what one would see in a spectrum obtained with traditional spectroscopic techniques. A close comparison with the average spectrum (Fig.~\ref{fig3}) confirms this. Tomogram 1 is the image comparable with that of a classic central stellar bulge. Although the eigenspectrum looks like a standard spectrum, it is not; the scale is not associated to intensity.

Eigenvector 2 contributes 0.088 per cent of the variance and displays, in combination with its tomogram, a clear map of the rotation of the emission line gas in the FoV. It is also clear from its tomography that this eigenvector is uncorrelated with the stellar component.
 
Eigenvector 3 contributes 0.032 per cent of the variance. Its characteristic is that it displays correlations among features that can be associated to emission line transitions. It is quite surprising that features related to two kinds of emission lines are visible: narrow lines, associated with the [\mbox{O\,{\sc i}}], [\mbox{N\,{\sc ii}}] and [\mbox{S\,{\sc ii}}] species and, also, H$\alpha$. But there is also a feature associated to a broad H$\alpha$ component. This component is typical of Seyfert 1 (or LINER type 1) galaxies and is usually taken as a clear evidence for an AGN associated with a supermassive black hole. This is, therefore, an important discovery, which  has never been reported before, despite the fact that this is a nearby galaxy. The broad lines associated to such features are emitted in the Broad Line Region -- BLR -- while the other, narrower, lines are emitted in the Narrow Line Region -- NLR. Features associated to [\mbox{O\,{\sc i}}] lines are also present in eigenvector 3, as they are in $E_2$, however, they were not visible in $E_1$.

Eigenvector 4 and its respective tomogram (contributing 0.013 per cent of the variance) shows again a correlation among the narrow line features, but this time it is anti-correlated to the broad H$\alpha$. Notice that in both $E_3$ and $E_4$ the emission lines features are correlated with the continuum in a complementary way. Eigenvectors 5 and 6 show correlations between narrow line features but involving distinct line widths.

One could attempt to interpret all eigenvectors up to the limit of 1976, the number of wavelength pixels (properties) in the datacube analysed here. However it is clear that the eigenvalues become smaller as the relative noise level of the eigenvector increases. In the present case the eight principal components explain 99.8979 per cent of the variance (see Table~\ref{eigenvalue}) remaining 0.1021 per cent of the variance contained in the other 1968 eigenvectors. Eigenvectors 7 and 8 are the last to be shown here (Appendix A). $E_7$ is still dominated by broad H$\alpha$, but the noise level clearly becomes strong and competes with any signal after $E_8$ (see Fig~\ref{fig1}).

\begin{table}
 \centering
  \caption{Eigenvectors and corresponding eigenvalues for the first eight principal components.\label{eigenvalue}}
  \begin{tabular}{@{}ccc@{}}
  \hline
   Eigenvector   & Eigenvalue           & Accumulated fraction  \\
    $E_k$        & (\% of the variance) & (\% of the variance)  \\

\hline
$E_1$ & 99.7443 & 99.7443 \\
$E_2$ & 0.0883  & 99.8326 \\
$E_3$ & 0.0325  & 99.8651 \\
$E_4$ & 0.0129  & 99.8781 \\
$E_5$ & 0.0084  & 99.8864 \\
$E_6$ & 0.0048  & 99.8912 \\
$E_7$ & 0.0039  & 99.8952 \\
$E_8$ & 0.0027  & 99.8979 \\
\hline
\end{tabular}
\end{table}

\subsection{The Broad Line Region: location and spectrum}

As mentioned earlier, obtaining the eigenvectors and tomograms is an objective process that does not depend on choices made by the user. However, by handling eigenvectors and tomograms, one can express aspects that do depend on the user's desires and skills. We will explore such aspects in the following.

It is clear that NGC 4736 has an AGN and that this AGN has a BLR. The question now is how to enhance this feature. This is a relevant question not only in the study of the properties of this emitting region but also determining the location of the AGN and, thus, the location of a supermassive black hole. As already mentioned in section~\ref{enhancement}, enhancing a feature ``A'' can be done by attributing the feature factor $\Gamma_k(A)$ to each eigenvector $k$, thus reconstructing the datacube. This can be done in two ways: equations~\ref{13} and \ref{14} provide the intensity cube, $I'_{ij\lambda}(BLR)$; alternatively, by attributing the factor $N_k$ (equation~\ref{15}) to each tomogram, one can construct the datacubes normalized to unit variance,  $V'_{ij\lambda}(BLR)$ (see equation~\ref{16}). This was done with the feature factors from Table~\ref{enhvects}.

\begin{figure*}
\centerline{\includegraphics[width=0.7\textwidth]{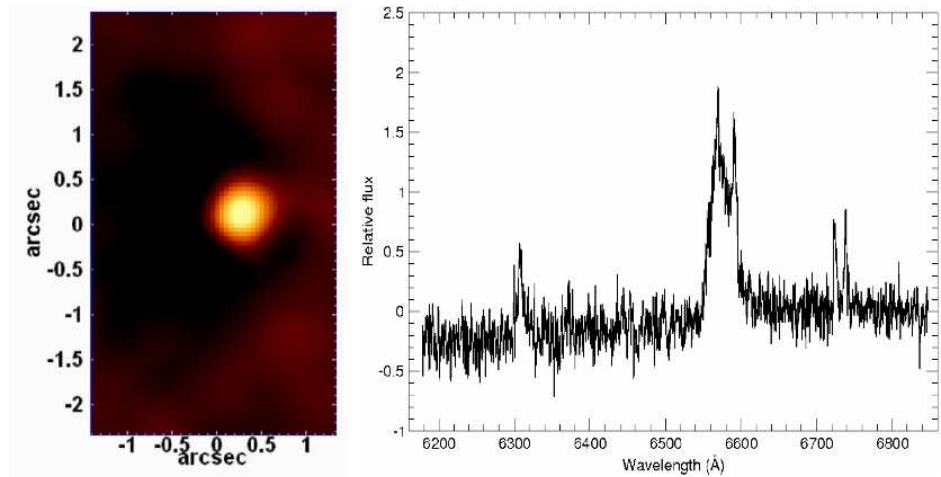}}
\caption{The image and spectrum of the BLR, extracted from the reconstructed datacube $I'_{ij\lambda}(BLR)$. The image is obtained from a ``narrow band'' (by adding consecutive images in wavelength pixels), centred on the red wing of (the broad) H$\alpha$; it, thus, maps the location of the BLR. The spectrum was extracted from a circle of 0.2 arcsec centred at the bright spot in the image. Notice that the broad H$\alpha$ emission is redshifted with respect to the rest frame (defined by the narrow lines). \label{fig4}}
\end{figure*}

\begin{figure*}
\centerline{\includegraphics[width=0.7\textwidth]{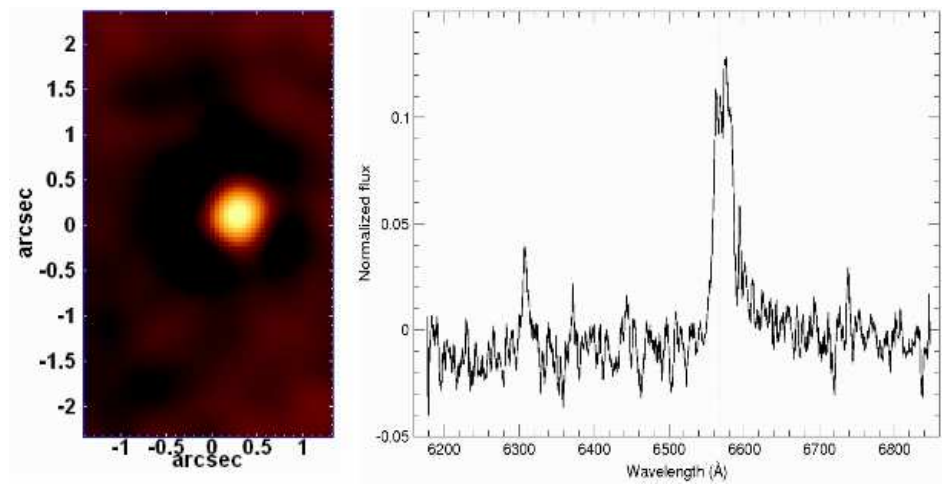}}
\caption{The image and the (smoothed) spectrum of the BLR, extracted from the reconstructed datacube $V'_{ij\lambda}(BLR)$ in a similar way as in Fig.~\ref{fig4}. In addition to the broad H$\alpha$, it is possible to see an asymmetric tail to the red. The vertical dashed line represents the wavelength of H$\alpha$ at the rest frame of NGC 4736.  \label{fig5}}
\end{figure*}

\begin{figure}
\centerline{\includegraphics[width=0.4\textwidth]{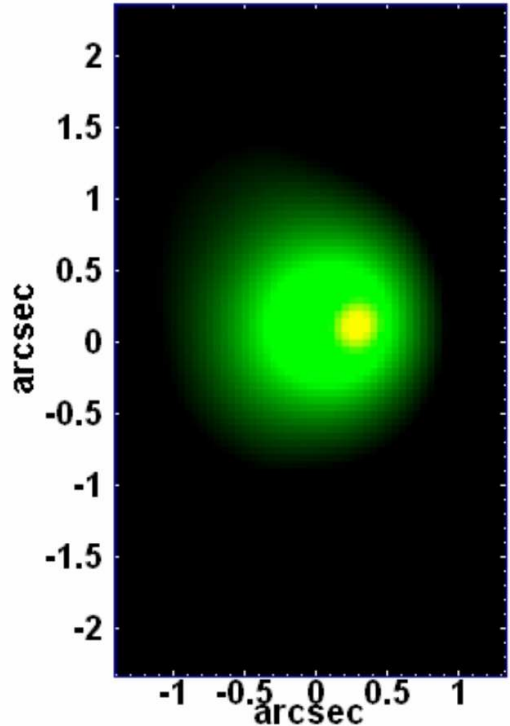}}
\caption{The location of the galactic stellar bulge (tomogram 1 is in green) and the BLR (image from Fig.~\ref{fig4} is in yellow) of the galaxy NGC 4736. The AGN is displaced by 0.15 arcsec from the centre of the galactic bulge. This corresponds to 3.5 parsec ($\sim10$ light years). Notice the bulge's comet-like shape.  These features are probably consequences of galaxy merger that occurred a few billion years ago \citep{steiner}. \label{fig6}}
\end{figure}

\begin{figure*}
\vspace{1.5cm}
\centerline{\includegraphics[width=0.7\textwidth]{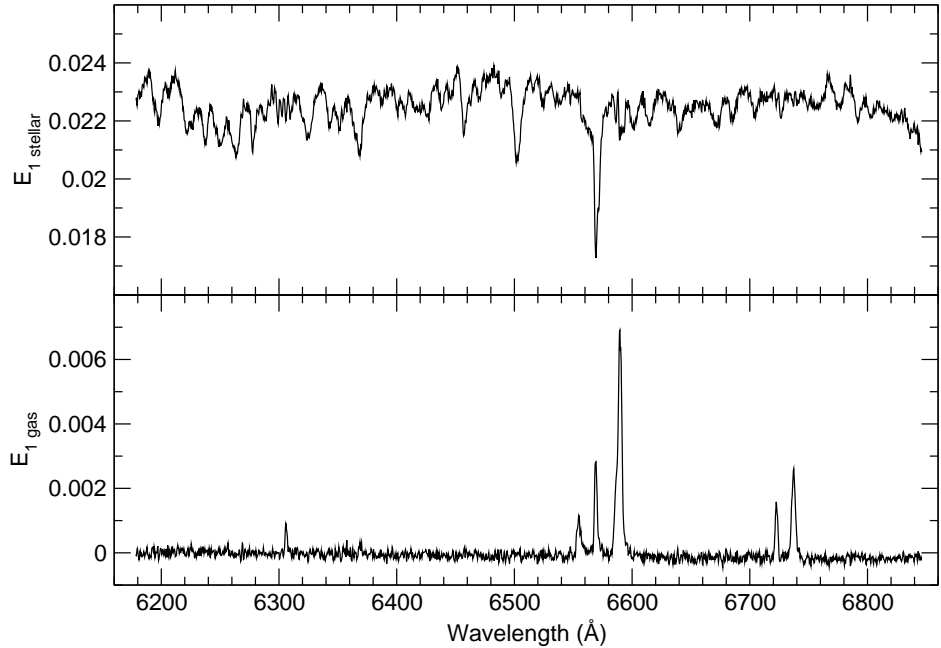}}
\caption{The stellar ($E_{1a}$) and gaseous ($E_{1b}$) components of eigenvector $E_1$; the sum of these two components is equal to $E_1$ and each of them can be obtained by adopting a template from the other.\label{fig7}}
\end{figure*}

\begin{figure}
\vspace{0.5cm}
\centerline{\includegraphics[width=0.5\textwidth]{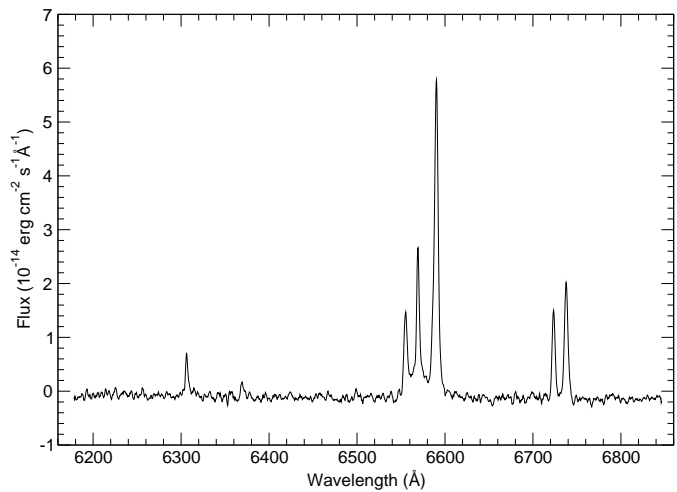}}
\caption{ The AGN flux calibrated spectrum of the LINER galaxy NGC 4736. \label{fig8}}
\end{figure}

From these reconstructed datacubes, we obtained the spectra and images of the BLR. The spectra of the BLR were extracted from a circular region centred on the AGN and with a radius of 0.2 arcsec. The BLR images were obtained with ``narrow filters'' obtained by adding consecutive images in wavelength pixels, centred on the red wing of the broad H$\alpha$ feature (Figs.~\ref{fig4} and \ref{fig5}).These images map the location of the BLR in the FoV and, thus, the position of the supermassive black hole.

\begin{table}
 \centering
  \caption{The feature factor for the BLR.\label{enhvects}}
  \begin{tabular}{@{}cc@{}}
  \hline
   Eigenvector   & $\Gamma_k(BLR)$   \\
\hline
$E_1$    & 0  \\
$E_2$    & 1  \\
$E_3$    & 1  \\
$E_4$    & 1  \\
$E_5$    & 0  \\
$E_6$    & 0  \\
$E_7$    & 1  \\
$E_8$    & 0  \\
\hline
\end{tabular}
\end{table}

By observing Figs.~\ref{fig4} and \ref{fig5}, one can note that, although presenting a lower signal-to-noise, the spectrum extracted from the cube $V'_{ij\lambda}(BLR)$ does a better job of separating the BLR from the NLR. This is as expected (see section~\ref{enhancement}) since all principal components enter with the same weight. As the Tomogram of the principal component 1 (see Appendix A) represents the image of the stellar bulge, one can, now, superpose the image of the BLR (Fig.~\ref{fig4}) onto the stellar component. This is shown in Fig.~\ref{fig6}
An interesting and surprising discovery is that the BLR, that locates the hypothetical supermassive black hole, is not positioned at the centre of the galactic bulge. This lack of positional coincidence is unexpected (to say the least) and certainly has important consequences for the study of this galaxy \citep{steiner}.

\subsection{The stellar and gas emitting components of eigenvector 1}\label{stellargas}

Eigenvector 1 is dominated by the correlation among the spectral properties of the bulge stars and gas emission. Eigenvectors of higher order are basically dominated by correlations of gas-emitting properties only. Could we create two datacubes, from eigenvector 1, the first representing the stellar continuum and the second, the gas line emission? To attempt this we proceeded in the following way: we took a NLR representative spectrum from the reconstructed datacube $I'_{ij\lambda}$ (this representative narrow emission line spectrum is shown in Fig.~\ref{fig7}) as template and scaled this template so to match the [\mbox{N\,{\sc ii}}] 6583 {\AA} line intensity in $E_1$. Subtracting this scaled template from $E_1$ leaves us with the stellar component of $E_1$. These two (stellar and gaseous) components of $E_1$ are shown in Fig.~\ref{fig7}. From these two vectors ($E_{1a}$ and $E_{1b}$) we reconstructed the respective datacubes using equation~\ref{11}. By adding now the $E_{1b}$ datacube to all others, from 2 to 10, we obtain the final cube containing the emission lines, $I'_{ij\lambda}(gas)$. We can now study the morphological and spectral properties of the line emitting gas.

\subsection{Flux calibration}\label{flux}

Flux calibration of a reconstructed datacube can be recovered by adding the average spectrum (as shown in equation~\ref{somaQlambda}). In general this process can only be done if the cube is reconstructed with all the components. Here, however, we neglected all eigenvectors above 10, as they essentially represent noise and add up to a tiny fraction of variance. But we are interested here in calibrating both datacubes of gas line emission and stellar component separately. To do this we proceed in the following way: using the spectrum of the gaseous component of $E_1$ (shown in Fig.~\ref{fig7}) as a template, we decomposed the average spectrum into its stellar and gaseous components in a way similar to the method used for eigenvector $E_1$ in the previous section. These two components are shown in Fig.~\ref{fig3} (the two narrow telluric lines seen in the upper diagram were removed ``by hand'' in the lower diagram). Then the stellar component of the average spectrum was added to the datacube obtained from the $E_{1a}$ component of eigenvector 1. Similarly, the gaseous component of the average spectrum was added to the datacube reconstructed from vector $E_{1b}$ and all others, from 2 to 10, obtaining now a flux calibrated gaseous cube $(I'_{ij\lambda}(gas))_0$. This decomposition of the average spectrum and their addition to the respective datacube can be done because the stellar and gaseous components add up linearly; so we end up having two flux calibrated datacubes in such a way that, if added together, they form the original calibrated datacube, except for the discarded noise.

\subsection{Extracting the AGN spectrum}

Finally we can extract the flux calibrated AGN spectrum from the flux calibrated cube $(I'_{ij\lambda}(gas))_0$. Notice that, both for constructing this cube and for calibrating it, PCA was crucial. As we know the location of the AGN (from Figs.~\ref{fig4} and \ref{fig5}), the extraction can be made taking a circular aperture of radius 0.5 arcsec (Fig~\ref{fig8}). The flux of the broad component of H$\alpha$ is $\sim 2.14 \times 10^{-13}$ erg s$^{-1}$ cm$^{-2}$. This corresponds to a luminosity of the broad H$\alpha$ component of $L \sim 6.14 \times 10^{38}$ erg s$^{-1}$. This luminosity is similar to that of NGC 4395, known currently as the least luminous Seyfert galaxy \citep{filippenko2}.

With a distance of 4.9 Mpc, this is one of the nearest type 1 AGN. Other objects with similar distances are M81, with a distance of 3.5 Mpc, NGC 4395 (4.1 Mpc) and Cen A (4.3 Mpc).

\section{Discussion and conclusions}

In this paper we presented the method of PCA Tomography and showed that it has differences and advantages when compared to traditional methods for analysing datacubes. With traditional spectroscopic techniques it would be difficult to show the existence of the BLR/AGN in NGC 4736; even more difficult would be to determine the position of the BLR with the accuracy we obtained here. The main results of this method can be summarized as follows:

\begin{enumerate}
\renewcommand{\theenumi}{(\arabic{enumi})}

\item PCA Tomography identifies eigenvectors, ordered in form of principal components according to the rank of the corresponding eigenvalues. Tomograms are images that represent ``slices'' of the data in the eigenvector’s space. The association of tomogram with eigenvectors is important for the interpretation of both. One can associate spectral characteristics to image features or \textit{vice-versa}.

\item One of the main advantages of PCA Tomography is the dimensional reduction. Instead of analysing tens of millions of pixels, one compresses the relevant information to a dozen of eigenvectors and tomograms that present these data in an organized fashion. This is also important for the data compression and transmission.

\item The fact that the eigenvectors are orthogonal among themselves is important for their handling and interpretation. When the datacube present uncorrelated physical phenomena, the orthogonality may be useful for identifying them. 

\item The reconstruction of the datacube with original format, but with separated (and eventually treated) components associated to distinct eigenvectors allows extracting spectra or images in order to isolate a given feature.

\item Besides, by selecting the eigenvectors or tomograms with certain correlations or anti-correlations, one can enhance features by reconstructing datacubes in original format with tomograms normalized to unit variance. This enhances the desired feature but may also increase the noise. 

\item Various types of noise may be eliminated or corrected by selecting their eigenvectors and tomograms: cosmic rays, hot/cold pixels etc. 

\item Flux calibration of the reconstructed datacubes is possible by adding the average spectrum. However, this is only possible directly when one takes into account all the components. In other situations, calibration might be possible but could be subtle. We illustrate this by applying the procedure to a specific case.

\end{enumerate}

In order to illustrate the PCA methodology we applied it to the central region of the LINER galaxy NGC 4736. The dimensional reduction of the data allowed the identification of characteristics that were unknown in advance. For example, we identify a type 1 nucleus, of very low luminosity, displaced from the centre of the stellar bulge. By handling the eigenvectors and tomograms we were able to display the spectra and locate the BLR (Figs.~\ref{fig4} and \ref{fig5}) of this AGN with respect to the galactic stellar bulge (Fig.~\ref{fig6}).

\vspace{1cm}
Those interested in software for PCA Tomography may obtain it on the PCA Tomography Homepage, at http://www.astro.iag.usp.br/$\sim$pcatomography.

\section*{Acknowledgments}
We would like to thank FAPESP -- Funda\c c\~ao de Amparo \`a Pesquisa do Estado de S\~ao Paulo -- for financial support under grants 06/05203-3 and 05/03323-9. We would also like to thank R. Cid Fernandes, L. Sodr\'e, Lucia Barroso and Keith Taylor for their careful reading of the manuscript and valuable suggestions. We are also indebted to the referee, Coryn Bailer-Jones, for many generous comments that allowed us to improve this paper.

\appendix
\section{Principal Components of NGC 4736}
Below we show the eigenvectors and tomograms of the 8 principal components of the nuclear region of LINER galaxy NGC 4736.

\vspace{1cm}

\begin{figure*}
\centerline{\includegraphics[width=0.7\textwidth]{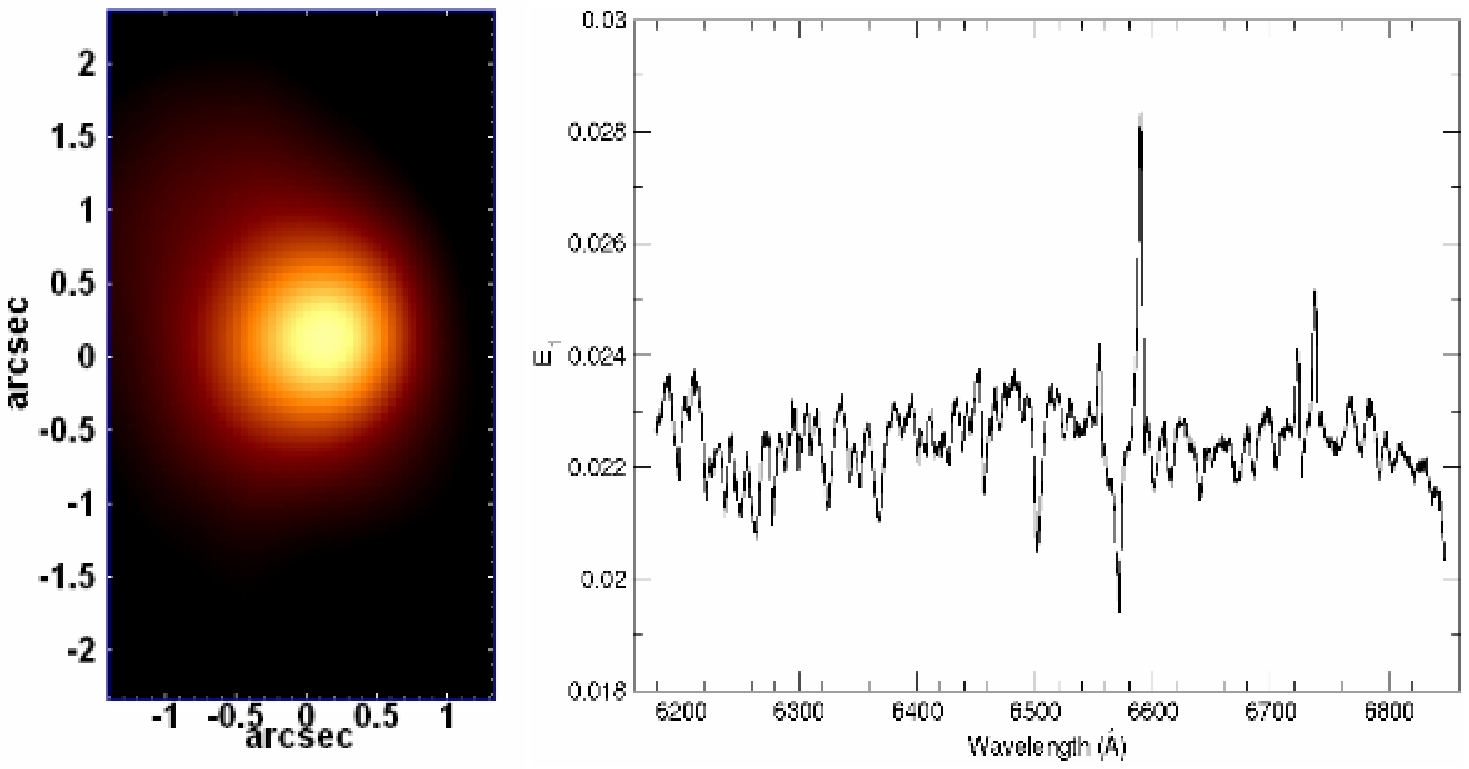}}
\caption{Tomogram of the principal component 1 and respective eigenspectrum.}
\end{figure*}

\begin{figure*}
\centerline{\includegraphics[width=0.7\textwidth]{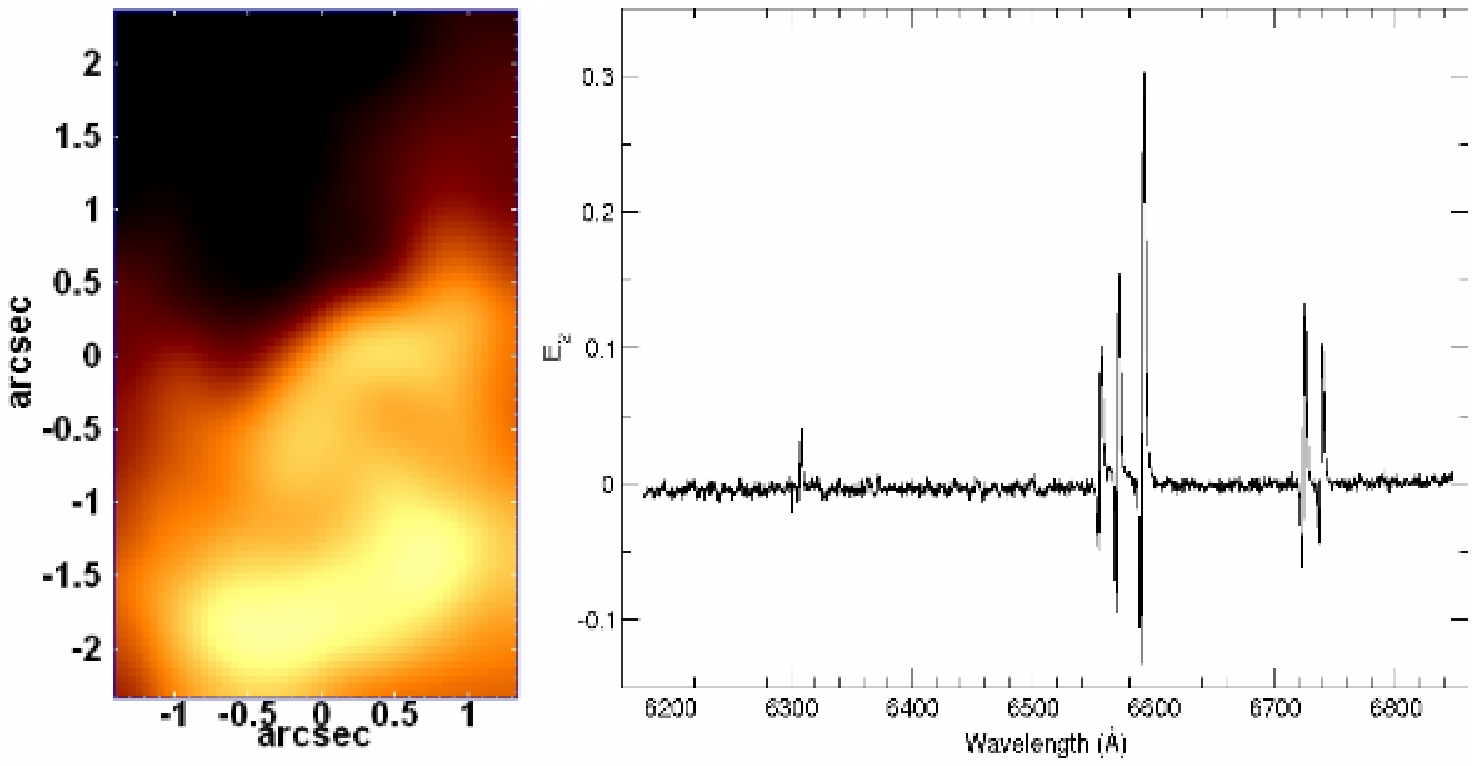}}
\caption{Tomogram of the principal component 2 and respective eigenspectrum.}
\end{figure*}

\begin{figure*}
\centerline{\includegraphics[width=0.7\textwidth]{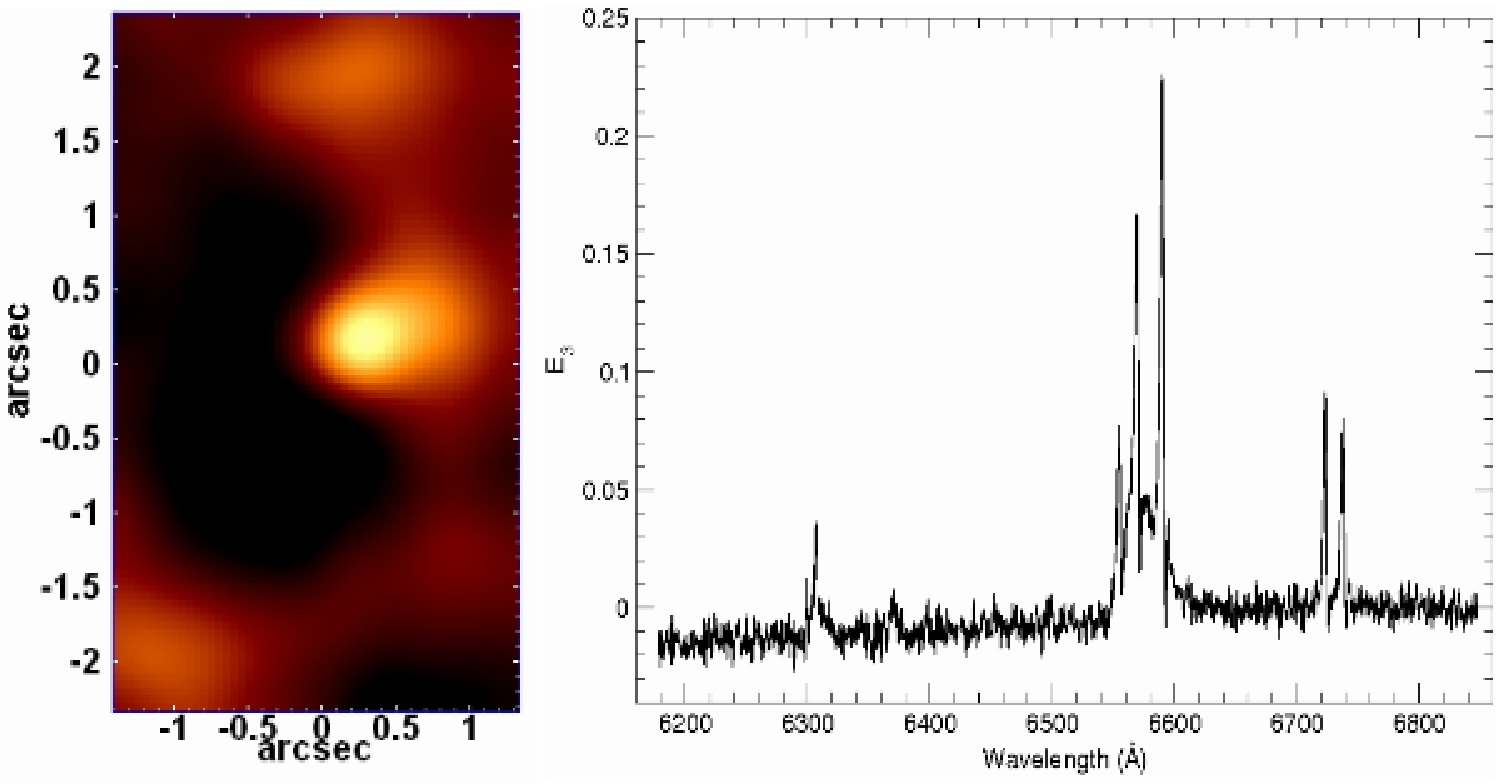}}
\caption{Tomogram of the principal component 3 and respective eigenspectrum.}
\end{figure*}

\begin{figure*}
\centerline{\includegraphics[width=0.7\textwidth]{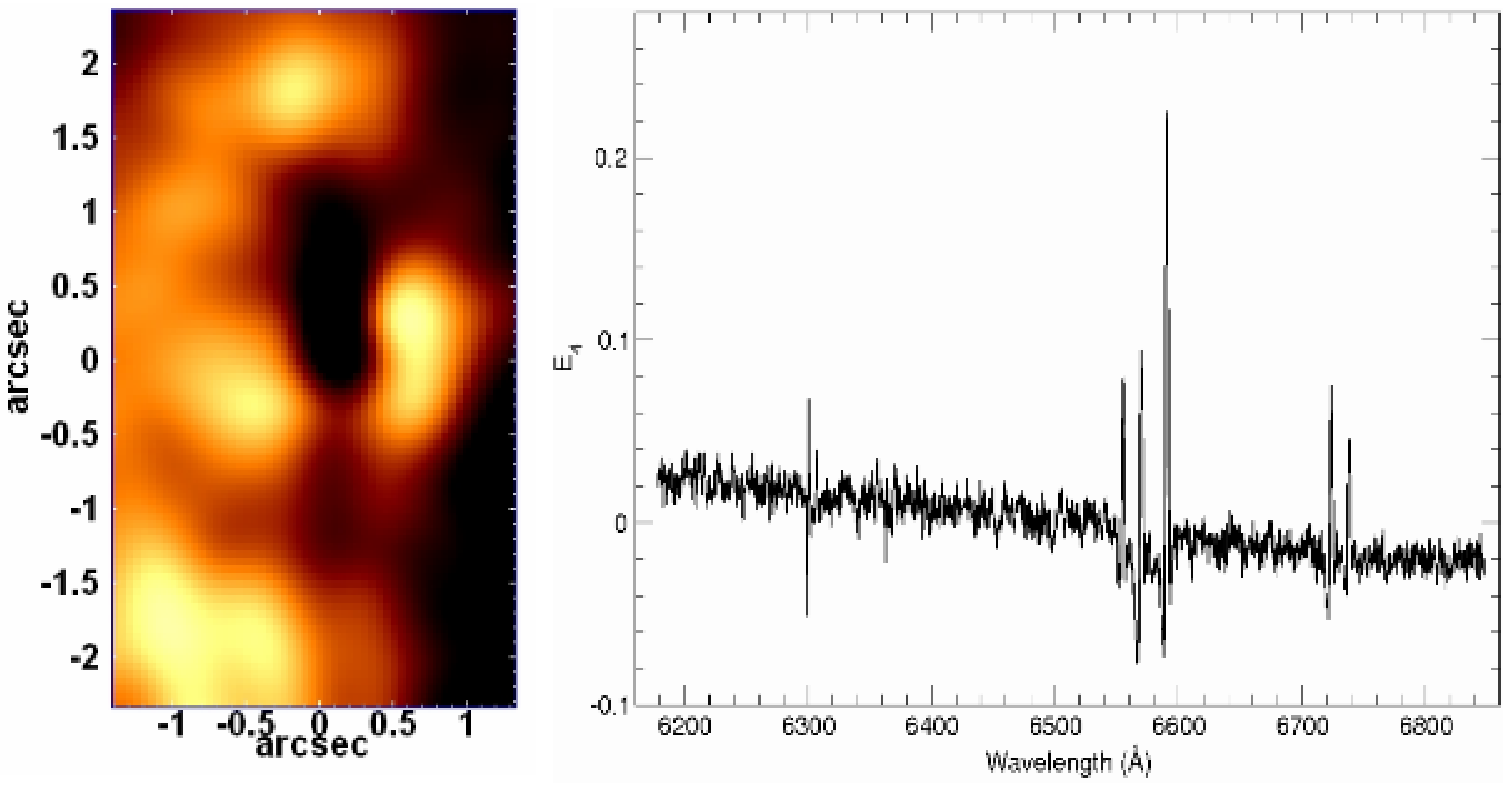}}
\caption{Tomogram of the principal component 4 and respective eigenspectrum.}
\end{figure*}

\begin{figure*}
\centerline{\includegraphics[width=0.7\textwidth]{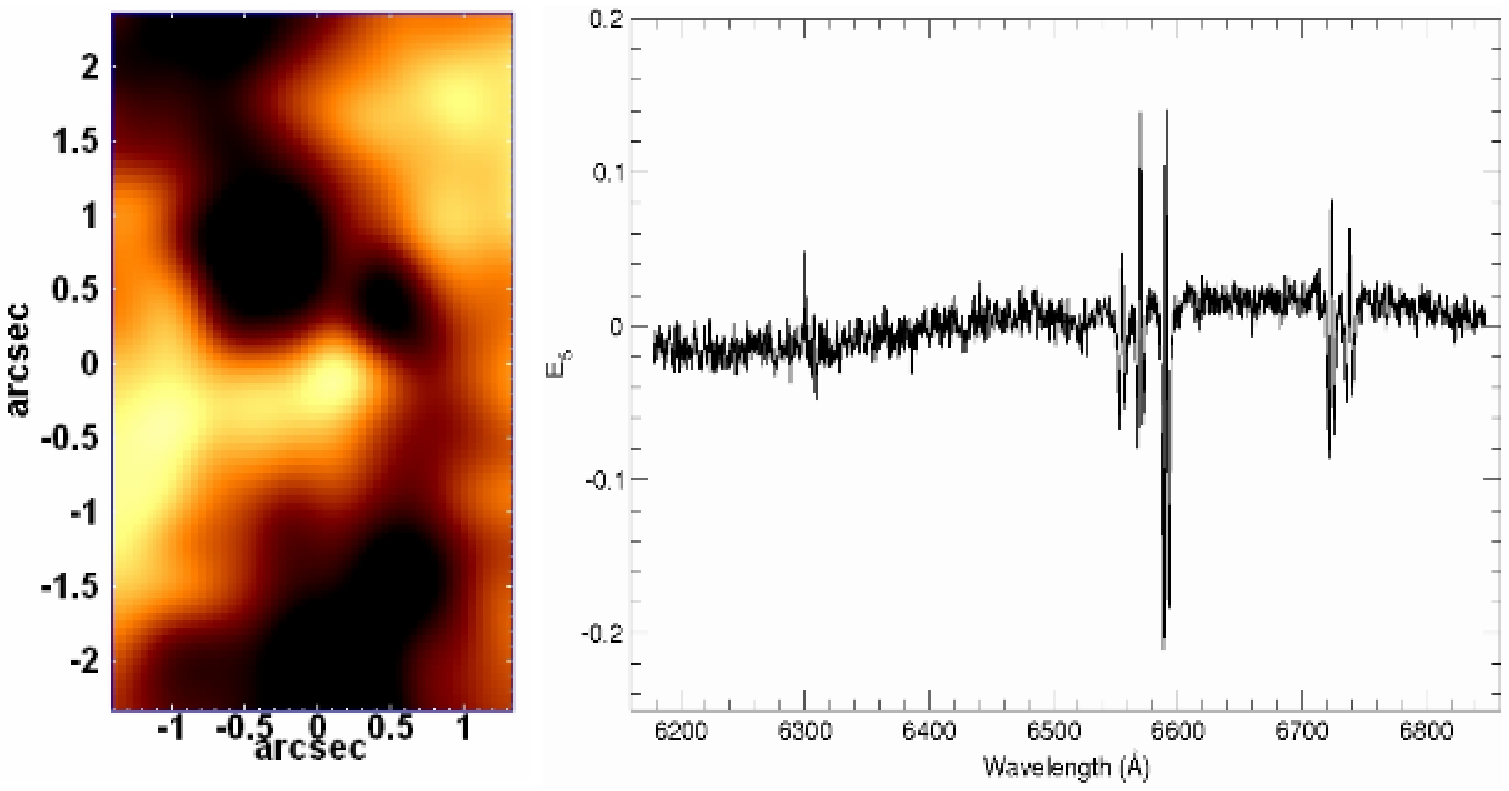}}
\caption{Tomogram of the principal component 5 and respective eigenspectrum.}
\end{figure*}

\begin{figure*}
\centerline{\includegraphics[width=0.7\textwidth]{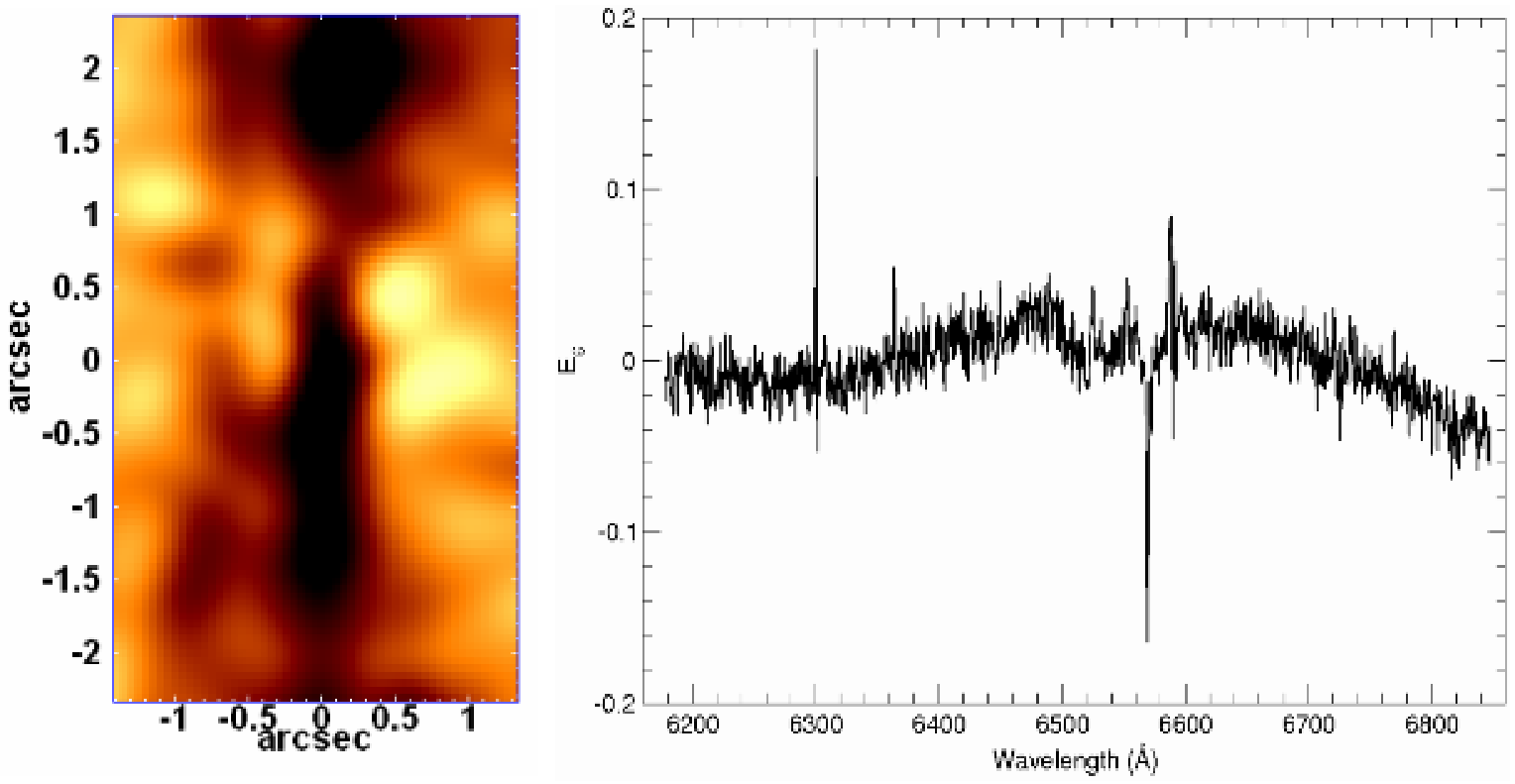}}
\caption{Tomogram of the principal component 6 and respective eigenspectrum.}
\end{figure*}

\begin{figure*}
\centerline{\includegraphics[width=0.7\textwidth]{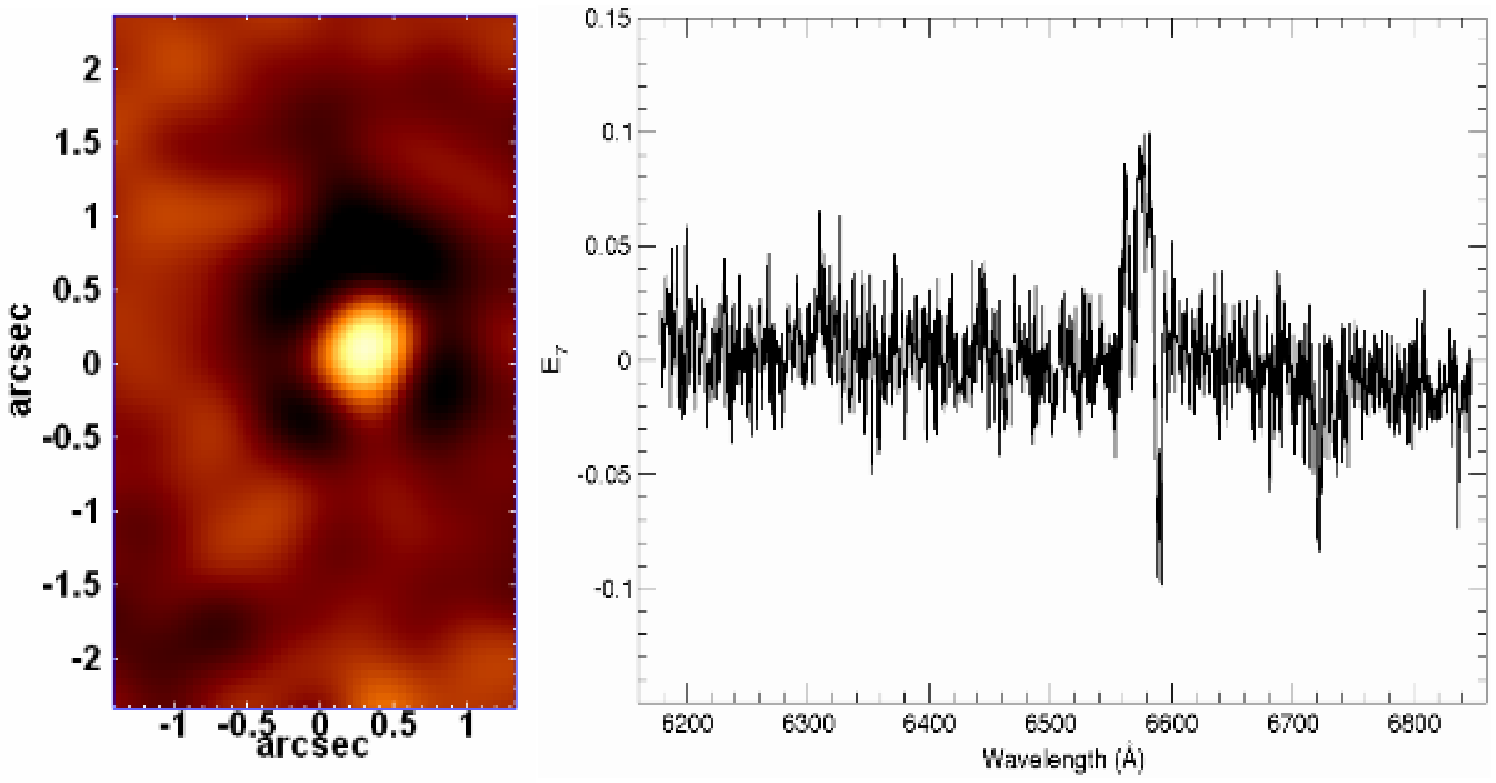}}
\caption{Tomogram of the principal component 7 and respective eigenspectrum.}
\end{figure*}

\begin{figure*}
\centerline{\includegraphics[width=0.7\textwidth]{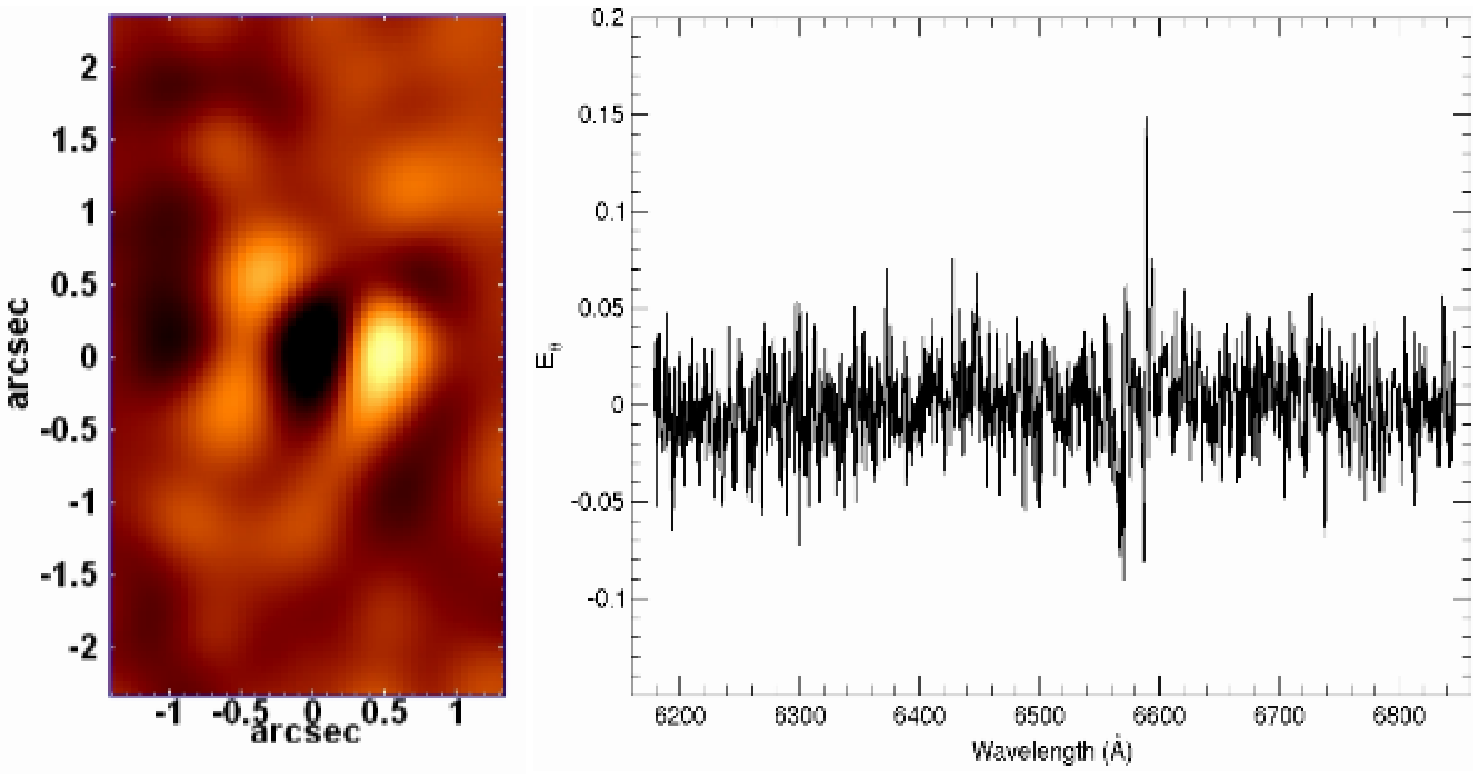}}
\caption{Tomogram of the principal component 8 and respective eigenspectrum.}
\end{figure*}

\label{lastpage}

\end{document}